\documentclass[conference]{IEEEtran}
\usepackage[linesnumbered,ruled,vlined]{algorithm2e}
\IEEEoverridecommandlockouts
\usepackage{amsmath,amssymb,amsfonts}
\usepackage{algorithmic}
\usepackage{graphicx}
\usepackage{CJKutf8}
\usepackage{makecell}
\usepackage{textcomp}\usepackage{bm}
\usepackage{bm}
\usepackage[numbers]{natbib}
\usepackage{xcolor}
\usepackage{hyperref}
\def\BibTeX{{\rm B\kern-.05em{\sc i\kern-.025em b}\kern-.08em
    T\kern-.1667em\lower.7ex\hbox{E}\kern-.125emX}}
\begin{document}

\title{ForPKG: A Framework for Constructing Forestry Policy Knowledge Graph and Application Analysis\\}

\author{\IEEEauthorblockN{Jingyun Sun, Zhongze Luo$^*$}
\IEEEauthorblockA{Northeast Forestry University, Harbin, China }
\IEEEauthorblockA{*Corresponding author, \textit{Member, IEEE}}
Email: sunjingyun@nefu.edu.cn, luozhongze0928@foxmail.com 
}

\maketitle
\begin{abstract}
A policy knowledge graph can provide decision support for tasks such as project compliance, policy analysis, and intelligent question answering, and can also serve as an external knowledge base to assist the reasoning process of related large language models. Although there have been many related works on knowledge graphs, there is currently a lack of research on the construction methods of policy knowledge graphs. This paper, focusing on the forestry field, designs a complete policy knowledge graph construction framework, including: firstly, proposing a fine-grained forestry policy domain ontology; then, proposing an unsupervised policy information extraction method, and finally, constructing a complete forestry policy knowledge graph. The experimental results show that the proposed ontology has good expressiveness and extensibility, and the policy information extraction method proposed in this paper achieves better results than other unsupervised methods. Furthermore, by analyzing the application of the knowledge graph in the retrieval-augmented-generation task of the large language models, the practical application value of the knowledge graph in the era of large language models is confirmed. The knowledge graph resource will be released on an open-source platform and can serve as the basic knowledge base for forestry policy-related intelligent systems. It can also be used for academic research. In addition, this study can provide reference and guidance for the construction of policy knowledge graphs in other fields. Our data is provided on Github \href{https://github.com/luozhongze/ForPKG}{https://github.com/luozhongze/ForPKG}.
\end{abstract}

\begin{IEEEkeywords}
Knowledge graph construction, large language models, government knowledge engineering, generative AI
\end{IEEEkeywords}

\section{Introduction}
Knowledge graph is a method of representing knowledge in graphical structure, where nodes represent entities or concepts in the real world, and edges represent relationships between entities \cite{wang2014knowledge}. Knowledge graphs provide an effective solution for managing, organizing, and utilizing structured data, and have widespread applications in search engines, recommendation systems, and intelligent question-answering, enabling more intelligent and personalized services for users \cite{chen2020review,fensel2020introduction}. Additionally, knowledge graphs are playing an important role in fields such as healthcare, finance, and the Internet of Things, providing support for decision-making and innovation in these areas \cite{li2024kgscs,ouyang2024modal}. Despite the fact that large language models (LLMs) can learn increasingly more prior semantic knowledge during pre-training, knowledge graphs remain an important knowledge representation pattern for assisting LLM inference, with their research popularity increasing rather than decreasing.

Most existing knowledge graphs are open-domain knowledge graphs, such as Dbpedia \cite{auer2007dbpedia}, CN-Dbpedia \cite{xu2017cn}, and YAGO \cite{suchanek2007yago}. These knowledge graphs typically depict common entities and their relationships in the real world, such as companies, singers, actors, and parent-child, spouse, and friend relationships. In addition, there are also some related studies on domain-specific knowledge graphs, such as maritime accident knowledge graph \cite{gan2023knowledge} etc. Policy knowledge graph is different from other domain-specific knowledge graphs, and its focus is not on depicting the relationships between real objects in the objective world, but more on depicting abstract knowledge such as concepts, terms, definitions, explanations, norms, and moral modes. A complete policy knowledge graph can assist in tasks such as compliance detection, intelligent policy question answering, intelligent policy retrieval, and government decision support, which has significant significance. However, to the best of our knowledge, there are not many related studies on policy knowledge graphs.

Existing policy knowledge graph research has only established relationships between policy documents and metadata, but lacks fine-grained semantic depictions of the internal text of policies. In contrast, this paper takes the forestry sector as its point of departure and designs a fine-grained ontology for forestry policy knowledge graph, not only migrating the common entities and relationship types from general knowledge graphs, but also proposing domain-specific entities and relationship types based on the characteristics of the forestry policy field and the theoretical system of moral logic. Moreover, knowledge extraction has always been a key process in constructing a knowledge graph, and existing knowledge extraction methods based on supervised models often require a large amount of labeled data. In the forestry policy field, there is less labeled data, and the construction cost is high. Therefore, this paper proposes a forestry policy knowledge extraction process based on open-source large language models. The process performs entity extraction, relationship classification, and tail entity recognition in sequence, thereby effectively identifying abstract relationships and long and difficult entity types in forestry policy texts.

The \underline{\textbf{For}}estry \underline{\textbf{P}}olicy \underline{\textbf{K}}nowledge \underline{\textbf{G}}raph \textbf{(ForPKG)} can serve as an important supplement to the knowledge graph of forestry and grassland field released by the Institute of Science and Information, China Forestry Science Institute, providing support and enrichment for forestry engineering compliance, forestry policy analysis, and forestry intelligent question answering tasks. In addition, the ontology and extraction process designed in this paper are not limited to the forestry field, but can be extended to other policy fields such as healthcare, insurance, and finance. Therefore, the methodology and practical experience provided in this paper can also serve as a reference and guidance for policy knowledge graph construction work in other fields.

In summary, the main contributions of this paper are as follows:
1) Propose a complete policy knowledge graph construction framework;
2) Release the first forestry policy knowledge graph resource;
3) Design the first fine-grained forestry policy knowledge graph ontology;
4) Propose a policy knowledge extraction process based on open-source large models;
5) Discuss the practical application of policy knowledge graphs on large model RAG.
\section{Related works}

\subsection{Knowledge graph}

Knowledge graphs represent various concepts, objects, and their interrelations in the real world through formal expressions of nodes (entities) and edges (relationships) \cite{peng2023knowledge,zhong2023comprehensive}. Based on their scope of application, they can be divided into open-domain knowledge graphs \cite{zirui2021survey} and domain-specific knowledge graphs \cite{abu2021domain}. Open-domain knowledge graphs mainly describe common entity types and their relationships in the real world, such as football players, actors, cities, and relationships like parent-child, spouse, and friendship. This type of knowledge graph is the most common and widely researched, examples include Dbpedia \cite{auer2007dbpedia} (currently the largest English open-domain knowledge graph), CN-Dbpedia \cite{xu2017cn} (the most frequently used open-domain Chinese knowledge graph), YAGO \cite{suchanek2007yago}, and WikiData \cite{vrandevcic2014wikidata}. In contrast to open domain knowledge graphs, the scope of a restricted domain knowledge graph is narrower, but it provides more in-depth and detailed information and knowledge within a specific field. Therefore, restricted domain knowledge graphs have a wide range of application scenarios for different field tasks \cite{chen2018automatic,fang2023knowledge,tang2023construction}. 

For the knowledge graph construction in policy and regulation domain, there are not many known related works, and the existing works are similar to the construction of literature network, only modeling the relationships between policy texts and various metadata, without delving into the internal semantics of policy texts for more fine-grained semantic relationships. However, in this study, the entire policy text was treated as a complete entity, thus hindering logical reasoning on the internal semantics of policies. To the best of our knowledge, there are no existing knowledge graph construction works that can simultaneously depict the external and internal semantic relationships of policy texts.

\subsection{Information extraction}

The construction of knowledge graph cannot be separated from information extraction, which involves extracting structured entities, relationships, and attributes from free text \cite{li2020real}. Traditional information extraction methods mostly model the entity extraction task as a sequence labeling problem and train a neural network model to predict the start and end positions of entity mentions in text \cite{liu2022chinese}. After identifying entity mentions in the text, it is possible to further predict the relationship categories based on them \cite{zeng2014relation}. Additionally, some methods extract entities and relationships jointly from text to avoid the error propagation from entity extraction to relationship recognition process \cite{chen2020joint}. Although these traditional information extraction methods have mature technical accumulation and in-depth research, these methods often require a large amount of labeled data. In certain specific domains or knowledge graph construction tasks for specific tasks, due to the lack of sufficient labeled data, these methods will be severely limited.

The emergence of large language models (LLMs) has made it possible for most natural language processing tasks to be performed with small or even zero samples. Information extraction tasks are no exception, and many researchers have proposed entity extraction and relationship recognition methods based on LLMs \cite{dagdelen2024structured,huang2023finbert}. Furthermore, many Chinese LLMs provide open-source APIs, so users can even input prompts and upload the text to be extracted to obtain extraction results. However, in the forestry policy knowledge graph ontology designed in this paper, there are many extremely long entity mentions. For example, the length of the abstract concept entity ``\textless \begin{CJK}{UTF8}{gbsn}退耕还林还草补助资金兑付\end{CJK} (Restore farmland to forests and grasslands, and provide subsidy funds.) \textgreater'' reached 12 characters in Chinese, while the length of the behavior entity ``\begin{CJK}{UTF8}{gbsn}对林业资金收支的真实、合法情况依法进行稽查监督。主要内容包括资金计划的申请、下达，资金的拨付、使用、管理及其他有关情况\end{CJK} (Lawfully audit and supervise the authenticity and legality of forestry funds' revenue and expenditure. The main content includes the application and issuance of funding plans, the allocation, usage, and management of funds, as well as other related matters.)'' reached 59 characters in Chinese. Moreover, compared with common entity relationships, the entity relationships designed in this paper's knowledge graph are more abstract. For example, the ``obligation'' relationship and the ``prohibition'' relationship. Therefore, how to effectively utilize the API of open-source LLMs to extract structured information from forestry policy and regulation texts and build a comprehensive forestry policy knowledge graph is still a topic that needs further research. However, to the best of our knowledge, there is no related work that can be referred to.

\section{Ontology design}\label{AA}

Ontology design is an important step in building a knowledge graph, especially for domain knowledge graphs. The design of the ontology directly determines the quality and application scope of the entire knowledge graph. In this paper, we first determine the entity types included in the forestry policy and regulation ontology, and then determine the relationship types and attributes.

\begin{table}[h]
\caption{The entity types of forestry policy regulation ontology defined in this paper.}
\label{tab:1}
\centering
\begin{tabular}{|>{\centering\arraybackslash}m{2cm}|c|>{\centering\arraybackslash}m{4cm}|}
\hline
Entity Types & Identifier & Description \\ \hline
Organizations & ORG & Companies, research institutions, government agencies, and non-profit organizations related to forestry \\ \hline
Person & PER & Individuals in the forestry field, including government officials, scientists, policymakers, environmental activists, and forestry entrepreneurs \\ \hline
Geographical Locations & LOC & General geographic locations or geographic entities related to forestry \\ \hline
Policy Documents & DOC & Policy document title \\ \hline
Categories & CLS & The forestry policy categories involved \\ \hline
Abstract Concepts & CONC & Terms, theories, and methods in forestry \\ \hline
Concrete Objects & OBJ & Specific tools, items, etc. related to forestry \\ \hline
Explanations/ Definitions & EXP\_DEF   & An explanation or definition of the concept of forestry \\ \hline 
Action & ACT & Actions that an able subject can perform \\ \hline
State & STATE & The state presented by an incapacitated subject \\ \hline
\end{tabular}
\end{table}

First, we borrow the most common entity types from general knowledge graphs and initially determine three entity types: organizations, person, and geographical locations. Then, we further determine two entity types related to the policy domain: policy documents and categories. Many previous works have also defined additional entity types such as cities, provinces, and issuing agencies, but we believe that these can all be classified into general entity categories such as geographical locations and organizations, so we do not repeat the definition. In addition, since there are many domain-specific concepts, domain-specific items, definitions, and explanations in forestry policy and regulation texts, we also set up three entity types of abstract concepts, concrete objects, and explanations/definitions. Finally, policy texts often contain a large number of instructions that are related to the ethical logic \cite{von1951deontic}, such as ``\begin{CJK}{UTF8}{gbsn}各级林业主管部门要进一步严把入口关，提高标准站建设成效\end{CJK} (the forestry authorities at all levels should further tighten the entry gate and enhance the effectiveness of standard station construction).'' . Therefore, this paper defines the two entity types of action and state based on the theoretical system of deontic logic. It is worth noting that, in addition to the two entity types defined based on the theoretical system of deontic logic, this paper also defines three relationship types, which will be elaborated on later. In summary, this paper defines a total of 10 entity types, and Table \ref{tab:1} shows all the defined entity types and their descriptions.

Based on the defined entity types and the characteristics of forestry policy and regulation field, this paper defines 15 relationship types: publish, locate, belong to, take office, have the duty, prohibit, have the right, define, be relevant to, be classified into, cite, contain, be published, employ, be cited. Among them, the obligation relationship, prohibition relationship, and authorization relationship are defined based on the theoretical framework of deontic logic \cite{von1951deontic}. Table \ref{tab:2} shows the defined relationship types and related information.

\begin{table}[h]
\caption{Relationship types defined in the forestry policy ontology.}
\label{tab:2}
\centering
\begin{tabular}{|>{\centering\arraybackslash}m{1.48cm}|>{\centering\arraybackslash}m{0.95cm}|>{\centering\arraybackslash}m{1.65cm}|>{\centering\arraybackslash}m{1.45cm}|>{\centering\arraybackslash}m{1.15cm}|}
\hline
Relationship Types & Identifier   & Domain of Definition                                                       & Domain                                                                     & Iinverse Relationship          \\ \hline
Publish            & publish      & ORG                                                                        & DOC                                                                        & Be Published (isPublished)     \\ \hline
Locate             & locate       & ORG/LOC                                                                    & LOC                                                                        & Contain (contain)              \\ \hline
Belong to          & belongTo     & ORG                                                                        & ORG                                                                        & Contain (contain)              \\ \hline
Take Office        & workFor      & PER                                                                        & ORG                                                                        & Employ (employ)                \\ \hline
Have the Duty      & duty         & PER/ORG/OBJ                                                                & ACT/STATE                                                                  & /                              \\ \hline
Prohibit           & isProhi-bited & PER/ORG/OBJ                                                                & ACT/STATE                                                                  & /                              \\ \hline
Have the Right     & hasRight     & PER/ORG/OBJ                                                                & ACT/STATE                                                                  & /                              \\ \hline
Define             & define       & CONC/OBJ                                                                   & EXP\_DEF                                                                   & /                              \\ \hline
Be Relevant to     & relevant     & \begin{tabular}[c]{@{}l@{}} CONC/OBJ/\\ EXP\_DEF/\\ ACT/STATE\end{tabular} & \begin{tabular}[c]{@{}l@{}}CONC/OBJ/\\ EXP\_DEF /\\ ACT/STATE\end{tabular} & This relationship is reflexive \\ \hline
Be Classified into & classify-To   & DOC                                                                        & CLS                                                                        & Contrain (contrain)            \\ \hline
Cite               & cite         & DOC                                                                        & DOC                                                                        & isCited (isCited)              \\ \hline
Contain            & contain      & \begin{tabular}[c]{@{}l@{}}DOC/LOC/\\ ORG/STATE\\ ACT/CLS\end{tabular}     & \begin{tabular}[c]{@{}l@{}}DOC/LOC\\ ORG/CONC\\ /OBJ\end{tabular}          & /                              \\ \hline
\end{tabular}
\end{table}

\section{Building process}

\begin{figure}[h]
    \centering
    \includegraphics[width=0.7\linewidth]{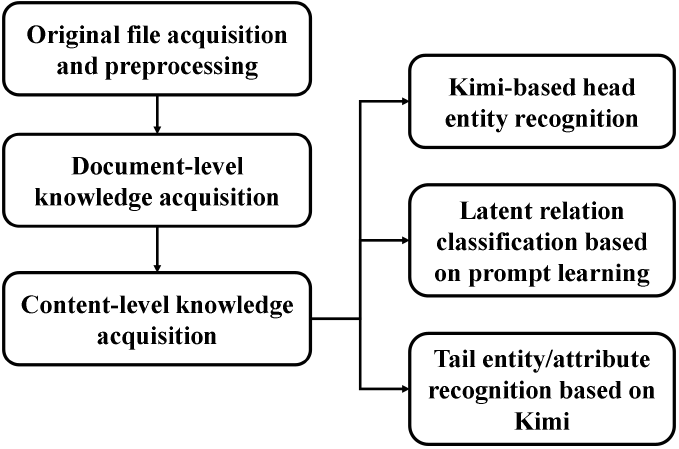}
    \caption{The workflow for building a forestry policy knowledge graph using an open-source LLM.}
    \label{1}
\end{figure}

Due to the lack of labeled data, existing knowledge graph construction methods based on supervised information extraction cannot effectively operate in specific domains. The emergence of large language models (LLMs) has effectively alleviated this situation. Many open-source Chinese LLMs allow direct API calls to achieve zero-shot condition information extraction, thereby assisting in constructing domain knowledge graphs. However, policy knowledge graphs have the following special points: 1) Some entities are mentioned very long, and may even occupy half of the entire sentence; 2) Some relationship types are more abstract, involving moral logic and other theories. Therefore, the common LLM calling schemes are not entirely suitable for this task. To solve this problem, this paper proposes a policy information extraction process based on open-source LLMs, as shown in Figure \ref{1}.

Specifically, the document-level knowledge acquisition process aims to acquire metadata such as the issuing agency and release time of forestry policy documents, as well as the category to which the policy documents belong and the similarity between policy documents. This process does not involve the internal knowledge of policy documents. The content-level knowledge acquisition process, on the other hand, delves into the text itself and extracts entities and their relationships from the text by calling the API of the LLM.

\subsection{Obtaining and preprocessing original documents}

All the original forestry policy and regulation documents used in this paper were obtained from the China Forestry Information Network\footnote{http://www.lknet.ac.cn/}, which was established and managed by the Forestry Science and Technology Information Institute of the Chinese Academy of Forestry, a leading authority in the forestry industry in China. This website is the largest and most comprehensive in the forestry industry in China. In this study, we initially selected 577 forestry policy and regulation documents issued by the State Forestry Administration from 1998 to 2018, and 180 documents issued by the State Forestry and Grassland Administration from 2018 to 2024, for a total of 757 documents. These documents were downloaded in TXT format and then sorted, with their metadata preserved. The State Forestry and Grassland Administration was established in March 2018, and the State Forestry Administration was disbanded after the government reform.

\subsection{Document-level knowledge acquisition}

Document-level knowledge refers to knowledge that is not related to the content of the text, including the organization that issued the document, the date of release, the implementation date, keywords, timeliness, etc. This type of knowledge is relatively easy to obtain, and can be obtained directly from the HTML structure when retrieving policy documents from a website.

Additionally, in this stage, the correlation between any two policy documents is calculated, thus establishing a relevant relationship between the policy document (DOC) entities. Given a set of policy documents $\mathcal{D}=\{d_{1},d_{2},...,d_{N}\}$, this paper uses the Chinese checkpoint of the pretrained language model BigBird (Zaheer et al., 2020) to encode each policy document into a semantic vector, resulting in $\bm{V}=\{\vec{v}_{1},\vec{v}_{2},...,\vec{v}_{N}\}$. Then, calculate the cosine similarity between every pair of vectors. If the similarity value between policy documents $d_{i}$ and $d_{j}$ is greater than the set threshold $\lambda$, establish a reflexive relationship edge $\textless d_{i},r_{i,j}^{rel},d_{j} \textgreater $ labeled ``relevant'' between the policy document entities $d_{i}$ and $d_{j}$. That's why we use BigBird to represent the entire policy text, as it supports text inputs up to 8,000 tokens in length.

\subsection{Content-level knowledge acquisition}

Extracting content-level knowledge from forestry policy documents requires a model to have a deep understanding of the semantic meaning of the text. Chinese LLMs are large language models pre-trained on massive amounts of Chinese language data, thus possessing good semantic understanding capabilities. Common Chinese LLMs include ChatGLM \cite{du2021glm}, LLaMa\_Chinese \cite{touvron2023llama}, BaiChuan \cite{yang2023baichuan}, and Kimi\footnote{https://kimi.moonshot.cn/}, among others. In this paper, the models were evaluated based on their performance, ease of use, and cost, and Kimi was ultimately selected. The paper found that asking Kimi a single question in a single round and having it extract all \textless entity-relation-entity\textgreater triples directly from the entire policy text resulted in poor performance. Therefore, this paper designed a three-step extraction process: the first step is entity recognition based on Kimi; the second step is potential relationship classification based on Prompt learning; and the third step is entity recognition based on Kimi. The following sections provide detailed descriptions of each step.

\subsubsection{Head entity recognition based on Kimi}

According to the forestry policy ontology designed in this paper, there are many long entity mentions in policy texts, such as ``\begin{CJK}{UTF8}{gbsn}由地方政府负责人担任林长，负责森林资源的保护和管理（对林长制的解释）\end{CJK} (the local government leader serves as the forest superintendent, responsible for the protection and management of forest resources (an explanation of the forest superintendent system)).'' Kimi has difficulty extracting entities of this type. Through observation, this paper found that such long entity mentions are almost always tail entities, so it decided to first have Kimi identify short and simple head entities instead. Additionally, in order to provide Kimi with more contextual information to improve its recognition ability, this paper inputs the entire text rather than individual sentences.

\subsubsection{Latent relationship classification based on prompt learning}
\label{4.3.2}

This step aims to identify the latent relationship types within the text based on the given text and the included head entities. Various prompt templates were repeatedly tested, but none effectively enabled the Kimi model to identify predefined relationship types from the given forestry policy texts. Considering that this process can essentially be viewed as a text classification task, this paper instead uses prompt learning to train a local text classification model. The reason for adopting prompt learning technology is due to its ability to achieve good model fine-tuning results in few-shot scenarios \cite{schick2020exploiting}. As is well known, obtaining labeled data in specific domains requires expensive human labor and time costs. In this paper, 10 training samples were carefully constructed for each relationship category, totaling 150 samples. Then, these 150 training data were used to fine-tune the DeBERTa model using prompt learning, as detailed in Algorithm \ref{ag1}.

\begin{algorithm}
    \renewcommand{\thealgocf}{1}
    \SetAlgoLined 
	\caption{Fine-tuning of relationship classification model based on prompt learning}
    \label{ag1}
	\KwIn{Training set $Train=\left\{\left(s_1,h_1,y_1\right),\left(s_2,h_2,y_2\right),\ldots,\left(s_{150},h_{150},y_{150}\right)\right\}$, Initial model parameters $\theta_{init}$}
	\KwOut{Optimized model parameters $\theta_{opt}$}
	Initialize the prompt template: $T(\cdot)$\; 
 	Mapping of labels to label words is initialized: $\phi:\ \mathcal{Y}\rightarrow\mathcal{V}$\;
 	Initialization loss: $\mathcal{L}=0$\;
	\For{$epoch=1 \to epochs$}{
 	\For{each ($s_i$,$h_i$,$y_i$) in $Train$}{
		 $\ell=\log{P(\left[MASK\right]=\phi(y_i)/T(s_i,h_i))}$  //Calculate the mask loss\;
		  $\mathcal{L}+=\ell$\;
		}
		Update parameters: $\theta_{init}=\theta_{init}-\alpha\cdot\frac{\partial L}{\partial\theta_{init}}$\;
	}
        ${\theta_{opt}} \gets {\theta_{init}}$\;
	\textbf{return} ${\theta_{opt}}$\;
\end{algorithm}

In this approach, $s_{i}$ represents the text to be recognized, $h_{i}$ is the head entity contained within it, and $y_{i}$ represents the gold-standard relationship type. The label word (Verbalizer) is set to the name of the relationship category itself.

The trained model is denoted as $\mathcal{M}$, given a text $s$ containing potential relationships and its head entity $h$, the trained model $\mathcal{M}$ can predict the potential relationship categories in $s$, as shown in Equation $\hat{y}=\mathcal{M}(s,h)$.

\subsubsection{Tail entity recognition}

Through the above process, we can obtain the head entities and the type of relationships contained in a given text. The acquisition of the tail entities, however, is relatively easy. By observing, we found that most of the tail entities are located immediately after the relationship words. For example, in the sentence ``\begin{CJK}{UTF8}{gbsn}灌木一般系指高3米以下，没有明显主干、呈丛生状态的木本植物\end{CJK} (bushes generally refer to woody plants that are less than 3 meters in height and have no obvious trunk and grow in a clump)'', the relationship word is ``\begin{CJK}{UTF8}{gbsn}一般系指\end{CJK} (generally refer to)'', and the tail entity is ``woody plants that are less than 3 meters in height and have no obvious trunk and grow in a clump.'' Therefore, we first let Kimi recognize the relationship word. Given the head entity and the type of relationship, we drive Kimi's API to perform the relationship word recognition.

\section{Experiment and analysis}

This section analyzes the feasibility and advanced features of the proposed forestry policy knowledge graph construction framework, specifically including an analysis of ontologies and information extraction methods. In addition, we analyze the practical application effects of the knowledge graph by using a RAG based on a large model.

\subsection{Ontological analysis}

\begin{figure}[hb]
    \centering
    \includegraphics[width=0.8\linewidth]{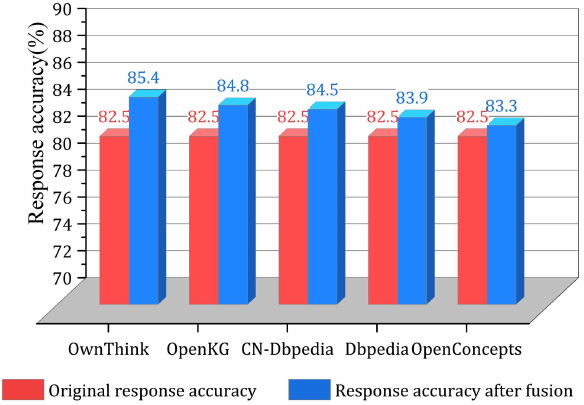}
    \caption{The improvement in response accuracy after merging the forestry policy knowledge graph with five general knowledge graphs.}
    \label{2}
\end{figure}

To verify the effectiveness and flexibility of the forestry policy ontology designed, we conducted the scalability experiment of the knowledge graph. We fused the forestry policy knowledge graph with five general knowledge graphs to test its query response ability after the fusion, thereby reflecting the flexibility and extensibility of the designed ontology. Figure \ref{2} shows the improvement in the accuracy of responding to user queries after fusing the forestry policy knowledge graph with OwnThink, OpenKG, CN-Dbpedia, DBpedia, and OpenConcepts. From the experimental results, it can be seen that the accuracy of responding to user queries has been significantly improved after fusing with the five general knowledge graphs. This shows that the ontology we designed can fully integrate with the ontology of general fields or other specific field knowledge graphs, thereby further improving its application effect in different real scenarios.

\subsection{Information Extraction Analysis}

\subsubsection{Overall Analysis}

This section provides a comprehensive analysis of the effectiveness and advanced nature of the policy information extraction method proposed based on open-source large models. To this end, we compare the proposed method with the following baselines: 1) DeepKE \cite{zhang2022deepke}, an open-source knowledge graph extraction framework, which supports low-resource and document-level information extraction; 2) LTP \cite{che2020n}, a Chinese processing foundation platform, which provides named entity recognition and relation classification functions; 3) Jiagu\footnote{https://github.com/ownthink/Jiagu} is an open-source entity and relation extraction tool; 4) DeepDive \cite{niu2012deepdive} is an open-source knowledge extraction system. OPENKG platform modified DeepDive to make it support Chinese. These baselines are off-the-shelf, unsupervised information extraction frameworks that represent the general level of unsupervised information extraction technology. Furthermore, considering that our method is based on the powerful language reasoning ability of LLM, we will also directly use large models to perform policy information extraction as a baseline for comparison, including: 1) Call Kimi's API directly to perform paragraph-level or document-level policy information extraction; 2) Use LLaMa2-Chinese directly for paragraph-level or document-level policy information extraction.

We evaluated the baseline and the method we proposed on the triad extraction from 50 forestry policy documents. Specifically, we manually annotated the 50 forestry policy documents to identify the triads contained in them, resulting in a total of 1,126 triads. Then, we calculated the precision and recall rates of the method on these 50 documents, and the final experimental results are shown in Table \ref{tab:3}.

\begin{table}[h]
\centering
\caption{Precision and recall of triple extraction.}
\label{tab:3}
\begin{tabular}{|lll|}
\hline
\multicolumn{1}{|l|}{\textbf{Method}}                            & \multicolumn{1}{l|}{\textbf{Precision}} & \textbf{Recall} \\ \hline
\multicolumn{3}{|l|}{A ready-to-use unsupervised information extraction framework}                                           \\ \hline
\multicolumn{1}{|l|}{DeepKE}                                     & \multicolumn{1}{l|}{22.3\%}             & 5.2\%           \\ \hline
\multicolumn{1}{|l|}{LTP}                                        & \multicolumn{1}{l|}{16.5\%}             & 2.2\%           \\ \hline
\multicolumn{1}{|l|}{Jiagu}                                      & \multicolumn{1}{l|}{32.9\%}             & 3.6\%           \\ \hline
\multicolumn{1}{|l|}{DeepDive}                                   & \multicolumn{1}{l|}{19.1\%}             & 1.9\%           \\ \hline
\multicolumn{3}{|l|}{Calling the LLM API directly}                                                                           \\ \hline
\multicolumn{1}{|l|}{Kimi- paragraph-level extraction}           & \multicolumn{1}{l|}{43.2\%}             & 17.9\%          \\ \hline
\multicolumn{1}{|l|}{Kimi- document-level extraction}            & \multicolumn{1}{l|}{50.7\%}             & 23.3\%          \\ \hline
\multicolumn{1}{|l|}{LLaMa2-Chinese- paragraph-level extraction} & \multicolumn{1}{l|}{38.6\%}             & 12.8\%          \\ \hline
\multicolumn{1}{|l|}{LLaMa2-Chinese- document-level extraction}  & \multicolumn{1}{l|}{30.7\%}             & 9.8\%           \\ \hline
\multicolumn{1}{|l|}{Our method}                                 & \multicolumn{1}{l|}{\textbf{76.2\%}}    & \textbf{62.6\%} \\ \hline
\end{tabular}
\end{table}

From Table \ref{tab:3}, it can be seen that existing off-the-shelf information extraction frameworks perform poorly in the task of policy information extraction. There are two main reasons for this: First, policy information contains a large number of entities with long and complex names, such as state entities, action entities, and explanation entities, and sometimes these entity names can take up half of the sentence. This presents a challenge to conventional information extraction frameworks. Second, many entities and relationships in policy information have a significant semantic gap with the entities and relationships in the general domain, so a non-customized, general-purpose information extraction framework is difficult to align with the extraction target. We can also see that directly calling the LLM's API cannot achieve the extraction results as good as our method. This indicates that our method is more suitable for policy information extraction by leveraging the powerful semantic reasoning ability of LLM while better adapting to the characteristics of policy information extraction. It has good performance in extracting policy concepts, terms, and long entities and their special relationships. Moreover, although our method's performance far exceeds the baselines, it has not yet reached 80\%, which means that the difficulty of policy information extraction task is still high, and further in-depth exploration is still needed in the future.

\subsubsection{Detail analysis}

This section evaluates the effectiveness of the key steps in our proposed method one by one, including entity recognition based on Kimi and potential relationship classification based on prompt learning. To present the results more intuitively, we use a radar chart to depict the extraction performance of these steps in handling different entity types and relationship types.

\begin{figure}[h]
    \centering
    \includegraphics[width=1\linewidth]{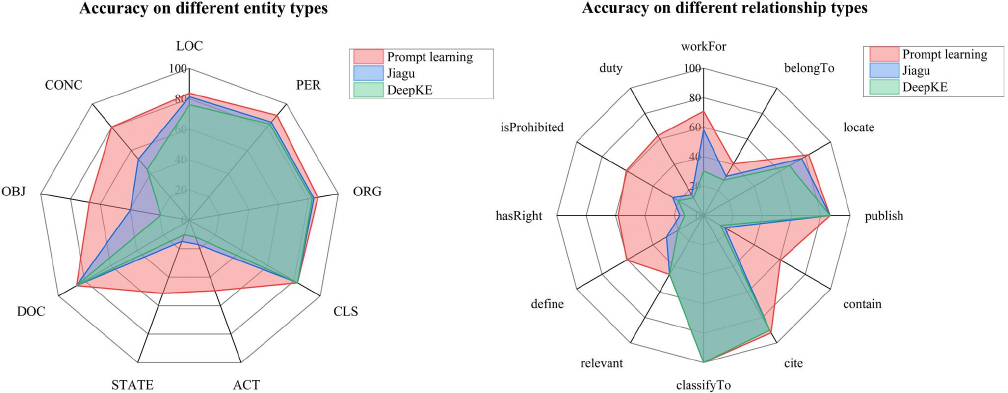}
    \caption{Accuracy on different entity types and relationship types.}
    \label{3}
\end{figure}

The left side of Figure \ref{3} presents the recognition accuracy rates of various entity types based on Kimi's entity recognition (blue areas). Observing the results, we can see that the recognition accuracy rates for ``action (ACT)'' and ``state (STATE)'' entities are relatively low, and the extraction results for ``explanation/definition (EXP\_DEF)'' entities are not satisfactory. These three types of entities are all long entities, which indicates that identifying long entities remains a major challenge in policy information extraction tasks. At the same time, we also notice that organizations, geographical locations, and person, which are consistent with general information extraction, are still the best recognized entities.

The right side of Figure \ref{3} shows the classification accuracy rates (blue areas) of the potential relationship classification based on Prompt learning for different relationship types. It is clearly visible from the figure that the accuracy rate of the ``relevant'' relationship type is the lowest. We suspect that this may be because the meaning of this relationship type is vague and the boundaries with other relationship types are not clear, leading to ambiguity in evaluating its effectiveness. In addition, the recognition performance of the ``belongTo'' relationship type is not satisfactory. We think this may be related to the low frequency of this relationship type in forestry policy texts, which makes it difficult to reflect the actual situation in the accuracy rate. The radar chart also shows the effects of the additional two baselines (red and green areas). Overall, it has shown satisfactory performance in the policy information extraction task, outperforming the other baselines, therefore, the policy information extraction we proposed is feasible and advanced.

\subsubsection{Data dependency analysis for training}

\begin{figure}[h]
    \centering
    \includegraphics[width=0.8\linewidth]{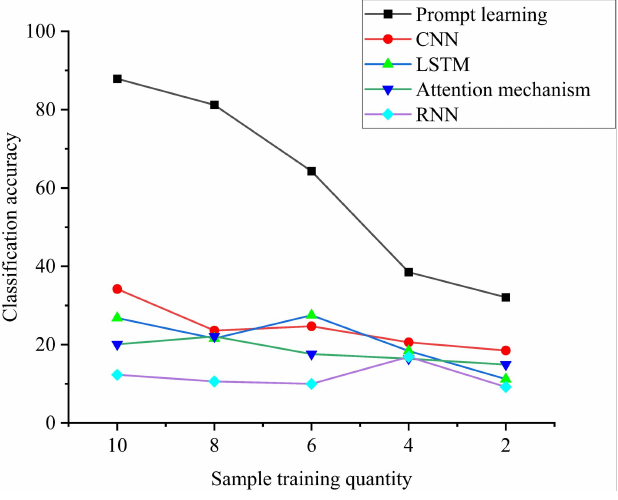}
    \caption{Training sample dependency analysis.}
    \label{4}
\end{figure}

We are committed to automating the policy information extraction process to minimize manual intervention. However, in the phase of relationship classification based on Prompt learning, we still need to manually annotate some data. As described in Section \ref{4.3.2}, we carefully selected and designed 10 training examples for each relationship category, totaling 150 training examples. This section aims to explore whether the number of training examples can be reduced without affecting the classification effect in the relationship classification process based on Prompt learning. The experimental results, shown in Figure \ref{4}, reveal a clear trend: as the number of training examples decreases, Prompt learning's performance in the relationship classification task significantly deteriorates. This indicates that fewer than 10 training examples are not sufficient to ensure the quality of classification for each relationship category. Furthermore, by comparing with the current mainstream relationship classification technologies, we find that the relationship classification method based on Prompt learning has obvious advantages in the scenario of data sparsity.

\subsection{Application-based practical analysis}

\begin{figure}[h]
    \centering
    \includegraphics[width=0.7\linewidth]{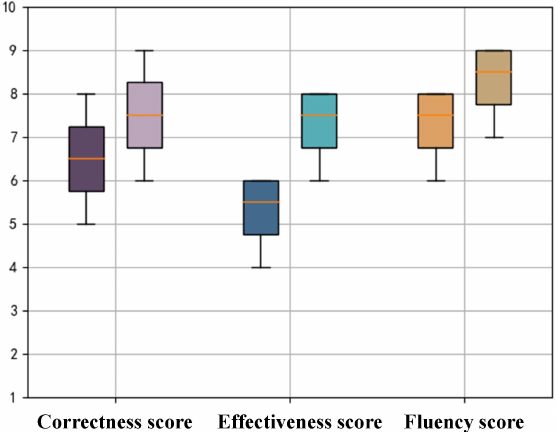}
    \caption{Evaluation results of answer generation quality before and after integrating the LLaMa-Chinese model with the knowledge graph of forestry policies.}
    \label{5}
\end{figure}

To further analyze the practical application effects of the forestry policy knowledge graph we have built, we used it as the RAG knowledge source for a large model. We used the LangChain framework to build a RAG system based on LLaMa-Chinese, and used GPT-4 to evaluate the effect of accessing the forestry policy knowledge graph on the text generation performance of the large model. The evaluation criteria cover three dimensions: correctness, effectiveness, and fluency of the text, with a scoring range of 1 to 10. We carefully designed a template to guide GPT-4 in performing the evaluation task, aiming to ensure the accuracy and consistency of the evaluation process. Figure \ref{5} clearly shows that after the introduction of the forestry policy knowledge graph, the LLaMa-Chinese model's scores on all three evaluation dimensions have significantly improved.

\section{Presenting the build results}

\begin{figure}[h]
    \centering
    \includegraphics[width=1\linewidth]{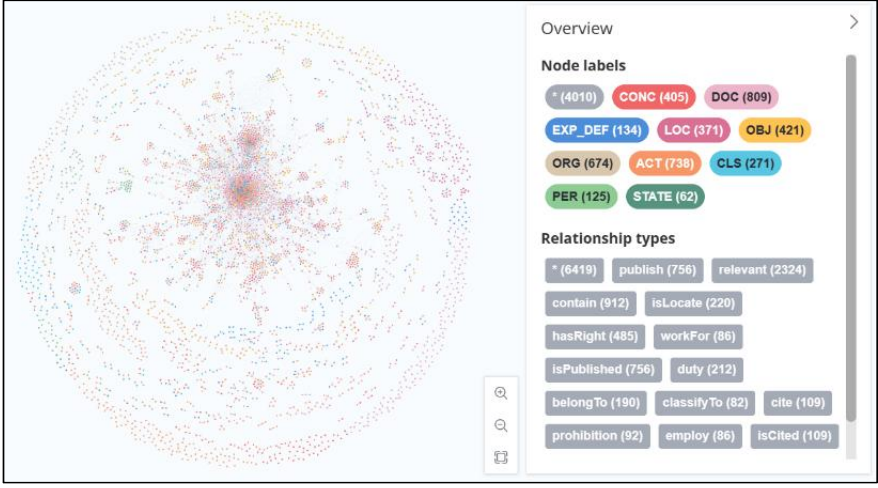}
    \caption{Visualization of the constructed forestry policy knowledge graph in a Neo4j database.}
    \label{6}
\end{figure}

We use the Neo4j database to store the entities and relationships of the forestry policy knowledge graph that we have built, which is visualized in Figure \ref{6}. In the future, we will further expand the scale of the knowledge graph and refine its ontology, eliminate noisy triples, and then package it to provide an online API for researchers in related fields to use.

\section{Conclusion}

This paper defines a fine-grained forestry policy knowledge graph ontology. Compared with existing work, the ontology defined in this paper delves deeper into policy content and has a finer-grained expressive ability. In addition, this paper proposes a three-stage policy knowledge extraction process based on open-source large-scale models and builds a fairly comprehensive forestry policy knowledge graph through the process. The final construction and analysis results show that the proposed policy knowledge extraction process is effective, and the forestry policy knowledge graph built has good expressive and reasoning capabilities for forestry policies.

\section*{Acknowledgment}

This research is funded by the Fundamental Research Funds for the Central Universities, with the project number 2572024BR31.

\bibliographystyle{IEEEtran}
\bibliography{name}

\end{document}